\documentstyle[aps,preprint]{revtex}
\newcommand{\be}{\begin{equation}}
\newcommand{\ee}{\end{equation}}
\newcommand{\ba}{\begin{eqnarray}}
\newcommand{\ea}{\end{eqnarray}}
\begin{document}
 
\begin{center}
{\Large \bf  Fokker-Planck equation of distributions of financial returns and power laws}
 
\vskip .17in
Didier Sornette$^{1,2}$
 
{\it $^1$ Laboratoire de Physique de la Mati\`ere Condens\'ee, CNRS
UMR6632\\ Universit\'e des Sciences, B.P. 70, Parc Valrose, 06108 Nice Cedex
2,
France \\
 
$^2$ Department of Earth and Space Science\\ and Institute of Geophysics
and
Planetary Physics\\ University of California, Los Angeles, California
90095, USA\\
 
}
\end{center}
 \vskip 3cm
\noindent
{\bf Abstract:} Our purpose is to relate the
Fokker-Planck formalism proposed by
[Friedrich et al., Phys. Rev. Lett. 84, 5224 (2000)] for the distribution of
stock market returns to the empirically well-established power law distribution
with an exponent in the range $3-5$. We show how to use Friedrich et al.'s
formalism to predict
that the distribution of returns is indeed asymptotically a power law with an 
exponent $\mu$ that can be determined from the Kramers-Moyal coefficients determined
by Friedrich et al. However, with their values determined for 
the U.S. dollar-German mark exchange rates,
the exponent $\mu$ predicted from their theory
 is found around $12$, in disagreement with the often-quoted
value between $3$ and $5$. This could be explained by the fact that the large 
asymptotic value of $12$ does not apply to real data that lie still far from
the stationary state of the Fokker-Planck description. Another possibility is
that power laws are inadequate.
The mechanism for the power law is 
based on the presence of multiplicative noise across time-scales, which is
different from the multiplicative noise at fixed time-scales implicit in the
ARCH models developed in the Finance literature. 

\pagebreak



\section{Summary of competing models for return distributions}

The distribution of asset returns provides the zeroth-order description 
of the statistical properties of asset returns. Over the years, several competing models 
have been proposed to describe the non-Gaussian heavy-tail structure of asset returns.
Without being exhaustive, we can cite:
\begin{itemize}
\item auto-regressive conditional heteroskedastic (ARCH) \cite{Engle} models which
assume that the variance is a function of past price variations 
and possibly of past variances (GARCH) \cite{Garch,Daco}; these models lead
in general to power tails;

\item truncated L\'evy laws \cite{Mantegna,bouchaud};

\item multiplicative noise models with reinjections \cite{Sorcont,Taka,Sorknopoff}
 which provide a mechanism for
 power law distributions \cite{deVries,Pagan,Guill,Gopikrishnan,Jondeau};
we note that such models are equivalent to ARCH models via a nonlinear change of variable
\cite{Haan,emmbre,Sorknopoff};

\item stretched exponential models \cite{Lahsor,Takasingapore} which can be derived
as tails of multiplicative processes \cite{Frisch};

\item multifractal cascade models from large to small time-scales \cite{Ghashghaie,Armusor,muzysor}.

\end{itemize}

In a recent letter \cite{Friedrich}, Friedrich et al. have proposed 
an alternative approach in terms of a Fokker-Planck
equation for the distribution of asset returns at different time-scales, relying on their
previous similar Fokker-Planck approach for the description of 
the distribution of velocity increments in turbulent cascades \cite{Fripeinke}
and their proposed analogy between turbulent cascades in hydrodynamics and information
cascades in stock markets.

Defining the price increment
\be
\Delta x(t,\Delta t) = x(t+\Delta t) - x(t)
\ee
over the time interval $\Delta t$ (which also defines the time-scale), 
the purpose of Friedrich et al. \cite{Friedrich}
is to write down an equation for the statistical process underlying the price
changes $\Delta x$ over a series of nested time 
delays $\Delta t_i$ (or time-scales) of decreasing durations. The controlling parameter is thus
$\Delta t$ while $t$ is dummy variable which, when varied, gives 
different realizations of
the same process at the given time-scale $\Delta t$. In other words, the problem
is to write down an equation of $\Delta x(\Delta t) $ as a function 
of time-scale $\Delta t$. In this goal, using the insight provided by cascade models,
it is natural to define the logarithmic time-scale
\be
\tau = \ln(t_0/\Delta t)~,
\ee
where $t_0=40 960s$ is the large time-scale from which the cascade proceeds.
Friedrich et al. \cite{Friedrich} then
arrive at the following Langevin equation for the price increment 
$\Delta x(\tau)$ as a function of logarithmic time-scale 
$\tau$
\be
d \Delta x(\tau) =  D^{(1)}(\Delta x) ~d\tau + \sqrt{D^{(2)}(\Delta x, \tau)}~dW~,
\label{nvbks}
\ee
where the Kramers-Moyal coefficients are given by
\ba
D^{(1)}(\Delta x) &=& -0.44~\Delta x~,\\
D^{(2)}(\Delta x, \tau) &=& 0.003 e^{-\tau/2} + 0.019 (\Delta x + 0.04)^2~,
\ea
for the U.S. dollar-German mark exchange rates. $dW$ is the
increment of the random walk $W(\tau)$, with white noise spectrum and variance 
equal to $2$. This Langevin equation (\ref{nvbks}) is equivalent to the
Fokker-Planck equation describing the distribution $P(\Delta x, \tau)$
of price variations $\Delta x$ at 
a given logarithmic time-scale $\tau$ \cite{Friedrich}:
\be
{\partial P(\Delta x, \tau) \over \partial \tau} =
- {\partial D^{(1)}(\Delta x) P(\Delta x, \tau) \over \partial \Delta x}
+ {\partial^2 D^{(2)}(\Delta x) P(\Delta x, \tau) \over \partial \Delta x^2}~. \label{ggkal}
\ee
For a fixed $\tau$, i.e., time-scale $\Delta t$, $P(\Delta x, \tau)$ is nothing
but the usual distribution of price increments studied in many empirical works.
The innovation brought by the formulation (\ref{ggkal}) or its equivalent stochastic 
version (\ref{nvbks}) is to relate the distribution at one time-scale to that 
at other time-scales. This formulation expresses in a familiar framework (the Fokker-Planck
or Langevin equations) the cascade
models already discussed in Refs.\cite{Ghashghaie,Armusor,muzysor}.
Let us stress again that the ``time'' $\tau$ entering in (\ref{ggkal}) and (\ref{nvbks}) 
is a time-scale, not a time along which one follows the trajectory of the price or of
the price increments. Thus, equations (\ref{ggkal}) and (\ref{nvbks})  are not
equations of the dynamical evolution of prices. They are mathematical representations
of the cascade model \cite{Ghashghaie,Armusor,muzysor}. In particular, 
equation (\ref{nvbks}) is a stochastic equation describing how price
increments change with time-scale. For a single realization of the noise trajectory 
$W(t)$, it describes one particular cascade path of the price increments as a 
function of time-scale. Constructing an ensemble of noise trajectories flowing from the 
large time-scale $t_0$ down to some fixed time-scale $\Delta t$ will then 
provide an ensemble of price variations at this time-scale $\Delta t$ from which 
the distribution at this time-scale can be determined. We now turn to examine how
(\ref{nvbks})  can used precisely in this goal of calculating $p(\Delta x, \tau)$.

\section{Mapping of the Langevin formalism of Friedrich et al.
onto intermittently amplifying multiplicative noise}

\subsection{Multiplicative noise with reinjection}

In all the derivation below, we thus consider a logarithmic time-scale $\tau$
flowing from $0$ (large time-scale $\Delta t=t_0$) to a fixed value $\tau$ 
corresponding to a fixed time-scale $\Delta t$. Our goal is to consider many
trajectories of equal logarithmic time-scale duration $\tau$ and derive the
corresponding distribution of price increments $\Delta x$ at this fixed time scale
$\Delta$. 

To perform this mapping, the key remark is that,
for large $\Delta x$, $\sqrt{D^{(2)}(\Delta x, \tau)}$ reduces to approximately
$\sqrt{0.019}~ \Delta x$, which corresponds to a multiplicative noise. 
In constrast, 
when $\Delta x$ becomes small,  $\sqrt{D^{(2)}(\Delta x, \tau)}$ reduces to 
a function independent of $\Delta x$ and the multiplicative noise is
transformed into an 
additive noise. 

These two limiting behaviors, when taken together, lead
exactly to the mechanism of multiplicative noise with reinjection,
shown to generate power law distribution, discussed in Ref.\cite{Sorcont}.
Specifically, discretizing (\ref{nvbks}) by considering a small time-scale
increment $d\tau$,
we get
\be
\Delta x(\tau + d\tau) = a(\tau) \Delta x(\tau)~,~~~~~~{\rm for~large}~~\Delta x~,  \label{jfkak}
\ee
where
\be
a(\tau) = 1 - 0.44 d\tau + \sqrt{0.019}~dW~.  \label{amnakla}
\ee
Note that $a(\tau)$ becomes larger than $1$ when $- 0.44 d\tau + \sqrt{0.019}~dW$ is 
positive, corresponding to an amplification or growth of $\Delta x(\tau)$.
In constrast, for small $\Delta x$, the expression (\ref{jfkak}) is changed into
\be
\Delta x(\tau + d\tau) = a_0(\tau) \Delta x(\tau) + \sigma dW~,~~~~~
{\rm for~small}~~\Delta x~,  \label{jfakakaa}
\ee
where $a_0= 1 - 0.44 d\tau$ and
\be
\sigma = 0.003 e^{-\tau/2} + 0.019 (0.04)^2~.
\ee
Without the time-scale dependence of $\sigma$, 
expression (\ref{jfakakaa}) would be nothing 
but the standard mean-reversal equation (albeit in time-scale
rather than time) or Orstein-Uhlenbeck process.

\subsection{Properties of the fluctuations of $\Delta x(\tau)$}

The stochastic dynamics described by (\ref{jfkak}) and (\ref{jfakakaa}) is the following.
For most of the time-scales, the realizations of the random noise $dW$ are small and the 
multiplicative coefficient $a(\tau)$ is smaller than $1$. As a consequence, $\Delta x$ shrinks
and eventually the additive noise equation (\ref{jfakakaa}) takes over and ensures that
$\Delta x$ does not go to zero. However, due to the intrinsic stochastic nature of the 
noise $dW$, there will be random occurrences of $dW$ that make the multiplicative factor 
$a(\tau)$ larger than $1$. When this occurs, $\Delta x$ is amplified. If this occurs
over several successive time-scale steps, $\Delta x$ is exponentially amplified. This is 
the mechanism that leads to heavy tailed distribution, by intermittent multiplicative
amplification. Such amplification is bound to have a finite lifetime since 
the expectation of the multiplicative growth rate $\langle \ln a(\tau) \rangle 
=a_0= 1 - 0.44 d\tau$ is negative.

We can make this argument more precise by quantifying
the power law distribution $P(\Delta x)$
generated by the mechanism involving the ``fight between 
exponentials'' \cite{mybook}. As we have just said, 
large $\Delta x$ are generated by intermittent amplifications
resulting from the multiplication by several 
successive values of $a$ larger than one. We now give a mean-field
argument to derive the result that this process produces distributions with
a power law tail.

We present the argument using a discretization of the time-scale in ``unit'' steps $d\tau$.
Let us call $p_>$ the probability
that the multiplicative factor $a$ 
is found larger than $1$. The probability
to observe $n$ successive multiplicative factors $a$ larger than $1$ over 
$n$ successive time-scale steps $d\tau$ is
thus $p_>^n$. Let us call $a_>$ the average of $a$ conditionned on being larger than $1$:
$a_>$ is thus the typical value of the
 amplification factor. When $n$ successive multiplicative
factors occur with values larger than $1$, they typically lead to an amplification
of the amplitude of $\Delta x$
by $a_>^n$. Using the fact that the equation (\ref{jfakakaa}) ensures that the amplitude
of $\Delta x$ 
remains of the order of $\sigma$ when the 
multiplicative factors $a$ are less than $1$,
this shows that a value of $\Delta x$ of the order of $\Delta x \approx \sigma a_>^n$ occurs
with probability 
\be
p_>^n = \exp \left( n \ln p_> \right) \approx \exp \left( \ln p_> 
{\ln {\Delta x \over \sigma} \over \ln a_>} \right)= {1 \over (\Delta x/\sigma)^{\mu}}~\,
\label{nckak}
\ee
with $\mu = -\ln p_> / \ln a_>$, i.e. 
\be
p_> a_>^{\mu} =1~.   \label{glalaz}
\ee
  This last expression represents
a kind of mean field version of the exact solvability condition 
\cite{Kesten,Haan,Calan,Sorcont,Sorknopoff} 
\be
\langle a^{\mu} \rangle = 1~,  \label{cnajka}
\ee
determining the exponent $\mu$ of the power law tail of the distribution of $\Delta x$:
\be
P(\Delta x) \sim \Delta x^{-(1+\mu)}~.  \label{gnalal}
\ee
The power law distribution is thus the 
result of an exponentially small probability of creating an exponentially
large value. Expression (\ref{nckak}) does not provide a precise determination of the
exponent $\mu$, only an approximate one since we have used a kind of mean-field
argument in the definition of $a_>$. It however illuminates the physical mechanism 
for the power law, as resulting from a fight between exponentials \cite{mybook}.

The exact value of the exponent is instead determined by the non-mean field condition 
(\ref{cnajka}) which is derived as follows. In the presence of multiplicative noise, the 
distribution $P(\Delta x)$ is solution of the integral equation which, for 
large $\Delta x$, reads
\be
P(\Delta x) = \int {da \over a}~ P_a(a)~\int d\Delta x ~P(\Delta x/a)~,  \label{galna}
\ee
where $P_a(a)$ is the distribution of the multiplicative factors defined in (\ref{jfkak}).
Looking for a power law solution (\ref{gnalal}) put in (\ref{galna}) gives 
directly the equation (\ref{cnajka}) on the exponent $\mu$. Our somewhat heuristic
summary presented here is backed up by exact rigorous analysis \cite{Kesten}.

\subsection{Numerical evaluation}

From expression (\ref{amnakla}), we see that $a(\tau)$ is Gaussian. For small time-scale
increments $d\tau$, $\ln a(\tau) = - 0.44 d\tau + \sqrt{0.019}~dW$ and we do not need
to distinguish between normality and lognormality, and we have
\ba
\langle \ln a \rangle &=& - 0.44 d\tau\\
\langle (\ln a)^2 \rangle - \langle \ln a \rangle^2 &=& 2 \times 0.019~ d\tau,
\ea
where we have used the property of the Wiener process $\langle (dW)^2 = 2d\tau$ 
(we keep the same convention as in Friedrich et al. \cite{Friedrich}  of a variance 
equal to $2$). This allows us to use the exact 
solution of (\ref{cnajka}) for the exponent $\mu$ \cite{Sorcont}:
\be
\mu = - {\langle \ln a \rangle \over \langle (\ln a)^2 \rangle - \langle \ln a \rangle^2}~.
\ee
The $d\tau$ dependence cancels out between the numerator and denominator, and with the value
obtained for the UD dollar-German Mark exchange rates, we get the estimation $\mu \approx 11.6$.
This value is significantly larger that the value in the range $3-5$ often reported in the literature 
\cite{deVries,Pagan,Guill,Gopikrishnan,Jondeau} for assets others than exchange rates.
One possible explanation is that 
exchange rates are known to exhibit thinner tails that stocks. Another explanation is that
the parameters of the distributions within the Fokker-Planck formalism are far from their
stationary values and the value of the exponent in the range $3-5$ could be a 
non-asymptotic value. Another explanation is that the description by power laws, even if
it has a rather long history, is not adequate \cite{Lahsor}.

\section{Conclusion}

In summary, we have used the stochastic Langevin formalism associated with the Fokker-Planck
description offered by Friedrich et al. \cite{Friedrich} to show that, in their model,
 the asympotic tail of the distribution of price variations at a given
 time-scale is a power law with a large 
 exponent $\mu \approx 11.6$. We have stressed that the 
 stochastic Langevin equation or equivalently the Fokker-Planck formalism are 
 expressed in terms of an evoluation along the time-scale axis rather than 
 along the time arrow. The power law distribution obtained in this model thus 
 relies intrinsically in the assumed cascade from large time-scales to small
 time-scales and is thus different from the multiplicative models 
 that are equivalent to usual ARCH formulations \cite{Haan,emmbre,Sorknopoff}, 
 since the later are defined
 for a fixed time scale. 
 
 However, since the multiplicative cascade across time-scales of Friedrich et al.
  as well as 
 multiplicative noises at a fixed time scale can both produce fat tail
 power law distributions, the empirical observation of such fat tails and their
 adequate fits by this formalism \cite{Friedrich} is insufficient to prove the reality of the 
 cascade. In order to demonstrate the existence of a genuine cascade across scales,
 it is necessary to calculate in addition the
  correlation functions between different time-scales and at 
 varying time lags: the asymmetry of the correlation functions across time-scales reported in 
 \cite{Armusor,muzysor} is, in our opinion, a convincing way
 of proving the existence of the cascade. This
 was also noted independently by the Olsen group in Zurich for the foreign
 exchange market, who coined it
 the HARCH effect \cite{Olsen,Daco}: 
 the coarse-grained volatility predicts the fine-grained volatility better than
the other way around. They also found this effect for the implied forward rates
derived from Eurofutures contracts \cite{Ballocchi}.

\end{document}